\documentclass[aps,prl,twocolumn,superscriptaddress]{revtex4}

\usepackage{graphicx}

\begin{document}

\renewcommand{\vec}[1]{{\mathbf #1}}

\title{Attojoule calorimetry of mesoscopic superconducting loops}
\author{O. Bourgeois}
\email[]{Olivier.Bourgeois@grenoble.cnrs.fr}
\affiliation{P\^{o}le de Biothermique et de Nanocalorim\'{e}trie,
Centre de Recherches sur les Tr\`{e}s Basses Temp\'{e}ratures, CNRS and
Universit\'{e} Joseph Fourier, 38042 Grenoble, France}
\author{S.E. Skipetrov}
\affiliation{Laboratoire de Physique et Mod\'elisation des Milieux Condens\'es,\\
Maison des Magist\`{e}res, CNRS and Universit\'{e} Joseph Fourier, 38042
Grenoble, France}
\author{F. Ong}
\affiliation{P\^{o}le de Biothermique et de Nanocalorim\'{e}trie,
Centre de Recherches sur les Tr\`{e}s Basses Temp\'{e}ratures, CNRS and
Universit\'{e} Joseph Fourier, 38042 Grenoble, France}
\author{J. Chaussy}
\affiliation{P\^{o}le de Biothermique et de Nanocalorim\'{e}trie,
Centre de Recherches sur les Tr\`{e}s Basses Temp\'{e}ratures, CNRS and
Universit\'{e} Joseph Fourier, 38042 Grenoble, France}

\date{\today}

\begin{abstract}
We report the first experimental evidence of nontrivial thermal behavior of the
simplest mesoscopic system --- a superconducting loop. By measuring the specific heat $C$ of an
array of 450,000 noninteracting aluminum loops with very high accuracy
of $\sim$20 fJ/K, we show that
the loops go through a periodic sequence of phase transitions (with period of an integer number
of magnetic flux quanta)
as the magnetic flux threading each loop is increased. The transitions are
well described by the Ginzburg-Landau theory and are accompanied by discontinuities
of $C$ of only several thousands of Boltzmann constants $k_B$.
\end{abstract}

\pacs{}

\maketitle


Halfway between micro- and macroscopic worlds, mesoscopic systems are known to exhibit a
series of unique phenomena which disappear on smaller as well as on larger
scales (see Refs.\ \cite{alt91,imry02,fazio03} for review):
oscillations of transition temperature in thin superconducting cylinders \cite{little62},
magnetic flux quantization \cite{sharvin81},
magnetoresistance oscillations \cite{webb85},
persistent currents \cite{levy90}, etc. Study of these phenomena is becoming increasingly
important because mesoscopic, several nanometers in size, elements will be the base of
this century's electronics and are likely to revolutionize many areas of human activity,
such as, for example, medicine, biotechnology, and information
processing \cite{imry02,fazio03,nano04}. Although the electric transport in mesoscopic
systems has been mainly studied during the past 20 years,
the thermal transport has also attracted some attention very recently \cite{schwab00,roukes99}.
Yet, very little is known about the thermodynamic and thermal behavior of mesoscopic systems
(i.e., behavior of their entropy, specific heat, etc.). Meanwhile, thermodynamics of
nanostructured systems, new phase transitions intrinsic for them \cite{deo00,bezr95},
the energies needed for their heating, the heat released when the system
changes its state, will certainly be important in numerous future applications of
nanoelectronic devices \cite{nano04}.

In the present Letter we report highly sensitive specific heat measurements performed on
the simplest mesoscopic system --- a superconducting loop of size comparable to the
superconducting coherence length $\xi(T)$ in a magnetic field. We show that even this simple
system exhibits behavior that differs substantially from that observed in macroscopic
superconductors. More precisely, we observe multiple phase transitions between states
with different numbers of magnetic vortices in the increasing magnetic field, accompanied by
discontinuities of the specific heat as small as only few thousands of $k_B$.
These mesoscopic phase transitions are due to the entrance of superconducting vortices into
the sample and have been recently observed in superconducting loops similar to ours by
susceptibility measurements \cite{zhang97} and by Hall magnetometry \cite{peder01,vodo03}.
Similar phenomena have been demonstrated to exist in mesoscopic disks \cite{geim97,baelus01}.

\begin{figure}
\includegraphics[width=6.8cm,angle=0]{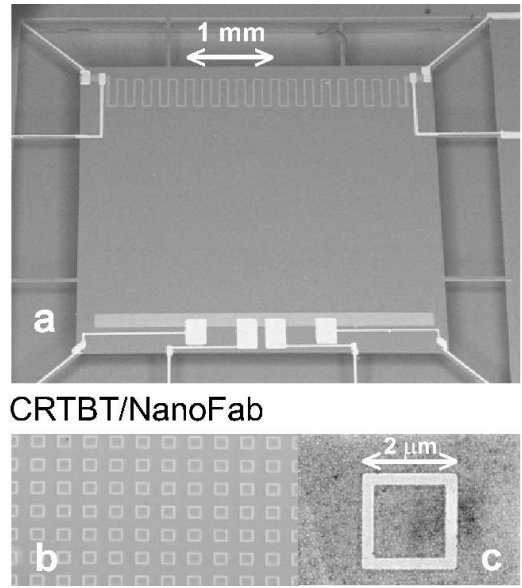}
\caption{\label{fig1}
(a). The suspended attojoule thermal sensor at 30$\times$ magnification.
The copper heater and the NbN thermometer can be distinguished in the upper and in
the lower parts of the silicon membrane, respectively.
The mesoscopic samples are deposited on the surface of the sensor between the heater and
the thermometer by e-beam lithography. (b). Scanning Electron Microscope (SEM) image of
the array of superconducting loops. (c). SEM image of a single aluminum superconducting loop.}
\end{figure}

Our sample is composed of 450 thousands nominally
identical, noninteracting \footnote{Although the inter-loop interaction can lead to
interesting phenomena when the mutual inductance of neighboring loops  is deliberately
maximized by placing the loops very close to each other \cite{dav97},
we do not expect it to affect our results significantly because 
for our sample the mutual inductance ($\simeq -20$ fH)
is much smaller than the
self inductance ($\simeq 5$ pH), the mutual magnetic flux always remains much smaller than $\Phi_0$,
and the energy of magnetic interaction between loops is much smaller than the free energy
of a single loop.} aluminum square loops (2 $\mu$m in size,
$w = 230$ nm arm width, $d = 40$ nm thickness, separation of neighboring loops $= 2$ $\mu$m, total
mass of the sample $m = 80$ ng),
patterned by electron beam lithography on a suspended sensor composed of a very thin (4 $\mu$m
thick) silicon membrane and two integrated transducers: a copper heater and a niobium nitride
thermometer \cite{fom97} (see Fig.\ \ref{fig1}).
The setup is cooled below the critical temperature $T_c$ of the superconducting transition by
a $^3$He cryostat and then its specific heat is measured by ac calorimetry. The technique of
ac calorimetry consists in supplying ac power to the heater, thus inducing temperature
oscillations of the thermally isolated membrane and thermometer \cite{fom97}.
For the operating frequency in the middle of the `adiabatic plateau' \cite{fom97}
(in our case, the frequency
of temperature modulation is $f \simeq 250$ Hz), the temperature of the system `sensor $+$ sample'
follows variations of the supplied power in a quasi-adiabatic manner, allowing measurements
of the specific heat $C$ with a resolution of $\delta C/C \gtrsim 5 \times 10^{-5}$
for signal integration times of the order of one minute. This makes it possible to measure
variations of the specific heat as small as 10 fJ/K (which corresponds to several thousands
of $k_B$ per loop), provided that the specific heat of the sensor (silicon membrane, heater,
and thermometer) is reduced to about 10--100 pJ/K. Since the temperature oscillates with
typical amplitude of few mK, our experimental
apparatus can detect energy exchanges of only few aJ at the lowest temperature of 0.6 K used
in our experiments.

\begin{figure}
\includegraphics[width=6.8cm,angle=0]{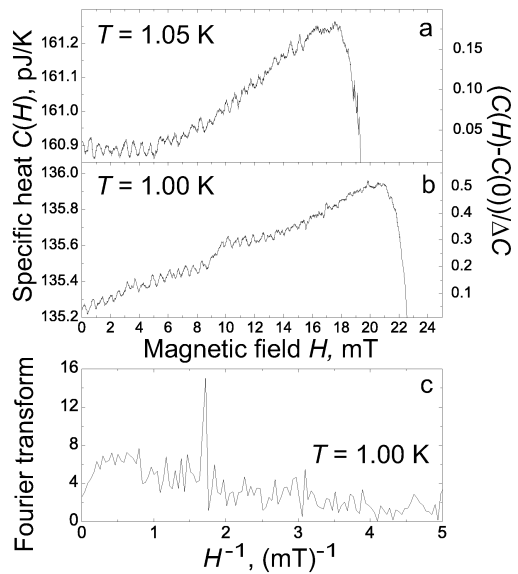}
\caption{\label{fig2}
Specific heat data at $T = 1.05$ K (a) and $T = 1.00$ K (b).
The right vertical axis shows the specific heat normalized by its jump $\Delta C$
at the critical field $H \sim 20$ mT where the transition to the normal state takes place
(only a part of the jump is shown in the figure).
For these measurements,
the current through the heater is 4 $\mu$A and the current through the thermometer is 0.1
$\mu$A. The amplitude of temperature oscillations measured at the thermometer is about
5 mK. When the magnetic field is varied, clearly seen oscillations of $C$ appear with
periodicity of 0.58 mT. This corresponds to the quantum of magnetic flux $\Phi_0$ threading a
square of 1.89 $\mu$m side. (c). Fourier transform of
the signal at $T = 1.00$ K. The peak at $1/H = 1.72$ mT$^{-1}$ corresponds to the periodicity
of 0.58 mT seen in the data.}
\end{figure}

In the absence of magnetic field, a discontinuity of the specific heat is observed at
$T_c = 1.2 \pm 0.02$ K, corresponding to the superconducting transition of the sample.
Once the sample is in the superconducting state, we keep the temperature constant
and turn on the magnetic field $\vec{H}$ directed perpendicular to the loops' planes. The field
is then varied very slowly (a typical run from 0 to 20 mT takes 10 hours),
and the specific heat
$C$ of the sample is measured for each value of $H$. Close to $T_c$ (above $T = 0.93 T_c$)
$C$ exhibits monotonic behavior with $H$ without any detectable regular fine structure,
whereas below $T = 0.93 T_c$ we observe oscillations of $C(H)$ with a period of approximately
0.58 mT (see Fig.\ \ref{fig2}).
The field $H \sim 20$ mT destroys superconductivity and produces a large jump $\Delta C$ of the
specific heat.
The periodic character of specific heat variations is even more
evident when looking at the Fourier transform of the data (see Fig.\ \ref{fig2}(c))
exhibiting a well-defined peak at $1/H \simeq 1.72$ mT$^{-1}$.
The observed period of oscillations of 0.58 mT corresponds to the magnetic flux quantum
$\Phi_0 = 2.07 \times 10^{-15}$ Wb through a square of 1.89 $\mu$m side, which is roughly
the size of our loops. At the same time, the amplitude of oscillations $\sim 50$ fJ/K
corresponds to specific heat variations of $\sim 0.1$ aJ/K ($\simeq 7500 k_B$) per loop.
A similar, periodic with the external magnetic field $H$, behavior of the magnetization $M$
of superconducting loops and disks has been recently reported in Refs.\ \cite{vodo03} and
\cite{geim97}. From these measurements, oscillations of $C$ versus $H$ can be anticipated by
using the Maxwell relation: $\partial C/\partial H \propto \partial^2 M/\partial T^2$,
suggesting that periodicity of $C$ with $H$ is a signature of entrance of vortices
into the mesoscopic sample.

\begin{figure}
\includegraphics[width=6.8cm,angle=0]{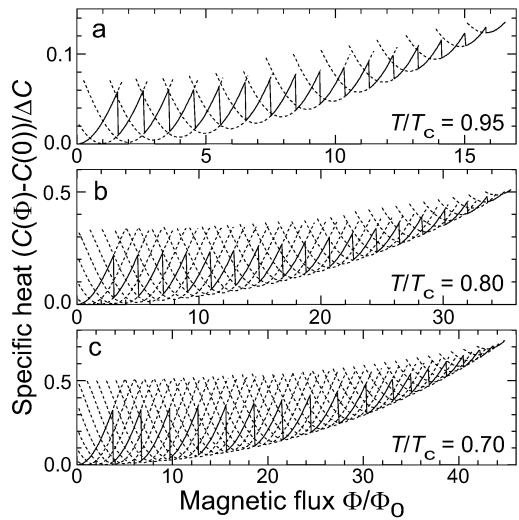}
\caption{\label{fig3}
Specific heat of a superconducting loop computed from the numerical solution of the
Ginzburg-Landau equation in the increasing magnetic field and for three different temperatures
(solid lines). Dashed lines correspond to states with constant vorticities
($n = 0, 1, 2, \ldots$ from left to right). $\Phi$ is the magnetic flux threading the loop,
$\Phi_0$ is the magnetic flux quantum, the curves are normalized by the discontinuity
$\Delta C$ of specific heat at the superconducting-to-normal transition that takes place
at $\Phi/\Phi_0 \simeq 16.5$ (a), $\Phi/\Phi_0 \simeq 35.5$ (b), and $\Phi/\Phi_0 \simeq 45.0$ (c).}
\end{figure}

A detailed description of the oscillations of $C$ with magnetic field can be obtained in the
framework of the Ginzburg-Landau (GL) theory of superconductivity \cite{tink04}.
We consider a circular loop of the same average perimeter as the actual
square loop and assume the GL coherence length at zero temperature $\xi(0) \simeq 0.15$ $\mu$m,
corresponding to the
mean free path $\ell \simeq 20$ nm measured independently. At a given magnetic flux
$\Phi$ threading the loop, the superconducting order parameter takes the form
$\psi(\vec{r}) = f_n(\rho) \exp(i n \phi)$, where the vorticity $n$ is a number of `giant'
magnetic vortices in the loop and we use cylindrical coordinates with the $z$ axis perpendicular
to the plane of the loop. The specific heat of the loop
$C = -T (\partial^2 F[\psi(\vec{r})]/\partial T^2)$, where $F[\psi(\vec{r})]$ is the GL
free-energy functional, is shown in Fig.\ \ref{fig3}(a) by dashed lines for different
values of vorticity $n$ and for $T = 0.95 T_c$. As follows from the figure, several values
of vorticity are possible at a given value of $\Phi$. In the thermodynamic equilibrium,
the transitions between the states with vorticities $n$ and $n + 1$
(so-called `giant vortex states') occur at $\Phi = (n + 1/2) \Phi_0$, which minimizes the
free energy and makes the specific heat to follow the lowest of curves corresponding to
different $n$ in Fig.\ \ref{fig3}. This results in oscillations of $C(\Phi)$ with a period of
$\Phi_0$, in agreement with the experiment. The amplitude of oscillations decreases when $T$
approaches $T_c$, and hence the oscillations are likely to be masked by noise at $T$
close to $T_c$, which explains why we did not succeed to observe any regular fine structure
in $C$ at $T > 0.93 T_c$.

It is, however, not clear if during the experiment the superconducting loops always remain in
the thermodynamic equilibrium as the magnetic field is varied.
Once the flux $\Phi$ surpasses $(n + 1/2) \Phi_0$, the giant vortex state $n$ becomes
metastable and, although a state with lower energy ($n + 1$ state) exists, the latter
is separated from the former by an energy barrier that can take a long time for the system to
overcome \cite{zhang97,baelus01}. As $\Phi$ is further increased, the height of the barrier
decreases and finally the barrier vanishes.
The state $n$ then becomes unstable and one more giant vortex enters into the loop.
To model this situation, we solve the GL equation for $\psi(\vec{r})$ by increasing the flux
$\Phi$ in small steps and using the previous-step solution as a starting point of our
relaxation-type algorithm. The resulting dependence of the specific heat on the flux exhibits
discontinuities at values of $\Phi$ where giant vortices enter into the loop.
The discontinuities are separated by $\Delta \Phi = \Phi_0$ and the curve $C(\Phi)$ has a
characteristic asymmetric triangular shape (solid line in Fig.\ \ref{fig3}(a)).
Since no unambiguous signature of such a shape is observed in the experimental data
of Fig.\ \ref{fig2}, we conclude that at not too low temperatures (at least, above $T = 0.9$ K)
our system always remains close to the thermodynamic equilibrium.
This is probably due to the electromagnetic and thermal noises which provide sufficient
energy for the system to overcome energy barriers between states with different vorticities
well before one of the states becomes unstable.

The maximum number of giant vortices that can be hosted by a loop of radius $R$ and arm width
$w \lesssim \xi(T)$ can be found to be 
$n_{max} \simeq \sqrt{3} R^2/w \xi(T)$, yielding the field $H_{max}$ at which oscillations of
$C$ should stop: $H_{max} \simeq \sqrt{3} \Phi_0/\pi w \xi(T)$.
Our experiments show that the loop can remain superconducting
up to $H$ slightly exceeding $H_{max}$, but the periodicity of $C$ with $H$ is lost beyond
$H_{max}$. We believe that this is due to the transition of the loop from the giant vortex state
for $H < H_{max}$ to the multivortex state for $H > H_{max}$, as discussed, e.g., in
Ref.\ \cite{baelus01}. At $H > H_{max}$, it becomes energetically favorable for vortices
(with a size of the order of $\xi(T)$) to enter into the width of the arms of the loop
instead of inside the loop itself and no periodicity of $C$ with $H$ is expected.

\begin{figure}
\includegraphics[width=6.8cm,angle=0]{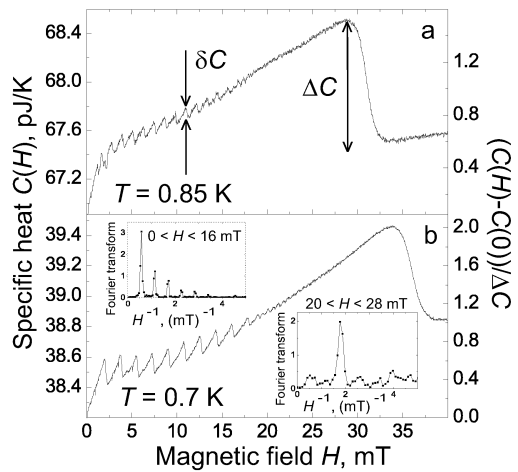}
\caption{\label{fig4}
Specific heat data at lower temperatures: $T = 0.85$ K (a) and $T = 0.7$ K (b).
The right vertical axis shows the specific heat normalized by its jump $\Delta C$
at the critical field where the transition to the normal state takes place.
At $T = 0.85$ K, the periodicity of $2 \times 0.58$ mT ($2 \Phi_0$ in terms of magnetic flux)
is observed,
whereas at $T = 0.7$ K (and at lower temperatures, down to $T = 0.6$ K) the periodicity of
$3 \times 0.58$ mT ($3 \Phi_0$) dominates. The insets show the Fourier
transforms of the data for $0 < H < 16$ mT and $20 < H < 28$ mT with dominant components
peaked at $1/H = 0.57$ mT$^{-1}$ and $1/H = 1.72$ mT$^{-1}$, respectively.
Note the existence of higher harmonics due to the triangular shape of the signal at low field.}
\end{figure}

To study the behavior of superconducting loops at lower temperatures (and hence at smaller
coherence lengths $\xi(T)$), we performed the measurements of the specific heat at $T = 0.85$ K
and $T = 0.7$ K (see Fig.\ \ref{fig4}). In contrast to Fig.\ \ref{fig2}, discontinuities of $C$
are now clearly seen, testifying that the loops do not remain in thermodynamic equilibrium
and explore metastable states. Another difference with Fig.\ \ref{fig2} is that the periodicity
of the signal is now a multiple of $\Phi_0$: $\Delta \Phi = 2 \Phi_0$ at $T = 0.85$ K and
$\Delta \Phi = 3 \Phi_0$ at $T = 0.7$ K. Such `multiple flux jumps' or `flux avalanches'
are signatures of simultaneous entrances of several (2 or 3) giant vortices into the loop
and can be considered as nonequilibrium phase transitions between states with $n$ and $m > n + 1$
giant vortices. Such transitions are also predicted by the GL theory (Fig.\ \ref{fig3}(b,c))
and have been recently observed in mesoscopic systems similar to ours by using the ballistic
Hall micro-magnetometry \cite{peder01,vodo03,geim97}.
Note that although the observed jumps of specific heat are rather sharp, their width
$\delta \Phi \sim 0.1 \Phi_0$ is finite. This is due to slightly fluctuating geometric parameters
of the loops, which therefore exhibit transitions between different giant vortex states at slightly
different values of the applied magnetic field. Mutual inductance effects are expected to
increase $\delta \Phi$ \cite{dav97}.
It is worthwhile to note that the discontinuities $\delta C$ of $C$ following from
the theory are systematically larger than the measured ones, but both the theory and the
experiment yield very similar results for $\delta C$ normalized by the large discontinuity
$\Delta C$ observed at the transition of the sample to the normal state (see Fig.\ \ref{fig4}
for the definitions of $\delta C$ and $\Delta C$).
$\delta C/\Delta C$ is typically several percent for $T$ close to $T_c$, but increases to
$\sim 25$\% for $T = 0.7$ K.

The data of Fig.\ \ref{fig4} have been obtained for the
magnetic field sweep rate of $2$ mT/hour, a ten times faster run yielded identical results.
However, for a ten times slower field sweep, finite lifetime of metastable states starts to play an
important role, leading to a more complicated behavior of $C(H)$
(alternating $\Phi_0$-, $2 \Phi_0$- and $3 \Phi_0$-jumps), which is a subject of ongoing study.
The multiple flux jumps can be understood by studying the stability of the state with vorticity
$n$ with respect to an admixture of a state with a different vorticity $m > n$. In the increasing
magnetic field, the instability first occurs for $m \simeq n + R/\sqrt{2} \xi(T)$,
favoring typical flux jumps $m - n = 1$ for $T$ close to $T_c$, when $\xi(T)$ is large
and $R/\xi(T) \sim 1$, and $m - n > 1$ at lower temperatures, when $\xi(T)$ decreases and
$R/\xi(T) > 1$. The multiple flux jumps are observed only in magnetic fields below
$H \simeq 18$ mT, whereas at larger fields and up to $H_{max}$ the single flux jumps are
recovered (see the insets of Fig.\ \ref{fig4}(b)), in agreement with magnetic measurements
\cite{peder01}. Theoretical curves also show the tendency to smaller periodicity
($\Delta \Phi = 2 \Phi_0$ instead of $\Delta \Phi = 3 \Phi_0$ in Fig.\ \ref{fig3}(c))
at high magnetic fields. 

In conclusion, the specific heat of mesoscopic superconducting loops is an oscillatory function of the
external magnetic field and exhibits discontinuities (jumps) at vortex entrance. Vortices enter
an isolated loop one, two, or three at a time, depending on the temperature at which the
experiment is performed. Definitely, the possibility of measuring the specific heat
of nanoscale objects with very high accuracy opens quite interesting prospects in the
field of meso- and nanoscale thermodynamics of superconducting as well as normal materials.

We would like to thank R. Maynard and G. Deutscher for fruitful discussions,
and P. Brosse-Maron, T. Fournier, Ph. Gandit, J.-L. Garden for the help.


\end{document}